\documentclass[a4paper,fleqn,usenatbib]{mnras}

\usepackage[T1]{fontenc}
\usepackage{ae,aecompl}
\usepackage{graphicx}	% Including figure files
\usepackage{amsmath}	% Advanced maths commands
\usepackage{amssymb}	% Extra maths symbols

\title[Fullerene/9-hydroxyfluorene cluster cations]
{Gas-phase formation of fullerene/9-hydroxyfluorene cluster cations}
\author[Wu et al.]{
	Yin Wu,$^{1}$
	Xiaoyi Hu,$^{2}$
	Junfeng Zhen,$^{3}$\thanks{E-mail: jfzhen@ustc.edu.cn}
	Xuejuan Yang,$^{1}$\thanks{E-mail: xjyang@xtu.edu.cn}
	\\
	$^{1}$ Hunan Key Laboratory for Stellar and Interstellar Physics, and School of Physics and Optoelectronics, \\
	Xiangtan University, Hunan 411105, China \\
	$^{2}$ Department of Chemical Physics, University of Science and Technology of China, 96 Jinzhai RD., Hefei, Anhui 230026, China\\	
	$^{3}$ Institute of Deep Space Sciences, Deep Space Exploration Laboratory, Hefei 230026, China\\
}

\begin{document}
	\label{firstpage}
	\pagerange{\pageref{firstpage}--\pageref{lastpage}}
	\maketitle

\begin{abstract}

In interstellar environment, fullerene species readily react with large molecules (e.g., PAHs and their derivatives) in the gas phase, which may be the formation route of carbon dust grains in space. In this work, the gas-phase ion-molecule collision reaction between fullerene cations (C$_{n}$$^+$, $n$$=$32, 34, ..., 60) and functionalized PAH molecules (9-hydroxyfluorene, C$_{13}$H$_{10}$O) are investigated both experimentally and theoretically. The experimental results show that fullerene/9-hydroxyfluorene cluster cations are efficiently formed, leading to a series of large fullerene/9-hydroxyfluorene cluster cations (e.g., [(C$_{13}$H$_{10}$O)C$_{60}$]$^+$, [(C$_{13}$H$_{10}$O)$_{3}$C$_{58}$]$^+$, and [(C$_{26}$H$_{18}$O)(C$_{13}$H$_{10}$O)$_{2}$C$_{48}$]$^+$). The binding energies and optimized structures of typical fullerene/9-hydroxyfluorene cluster cations were calculated. The bonding ability plays a decisive role in the cluster formation processes. The reaction surfaces, modes and combination reaction sites can result in different binding energies, which represent the relative chemical reactivity. Therefore, the geometry and composition of fullerene/9-hydroxyfluorene cluster cations are complicated. In addition, there is an enhanced chemical reactivity for smaller fullerene cations, which is mainly attributed to the newly formed deformed carbon rings (e.g., 7 C-ring). As part of the coevolution network of interstellar fullerene chemistry, our results suggest that ion-molecule collision reactions contribute to the formation of various fullerene/9-hydroxyfluorene cluster cations in the ISM, providing insights into different chemical reactivity caused by oxygenated functional groups (e.g., hydroxyl, OH, or ether, C-O-C) on the cluster formations.

\end{abstract}


\section{Introduction}

\begin{figure*}
	\centering
	\includegraphics[width=\textwidth]{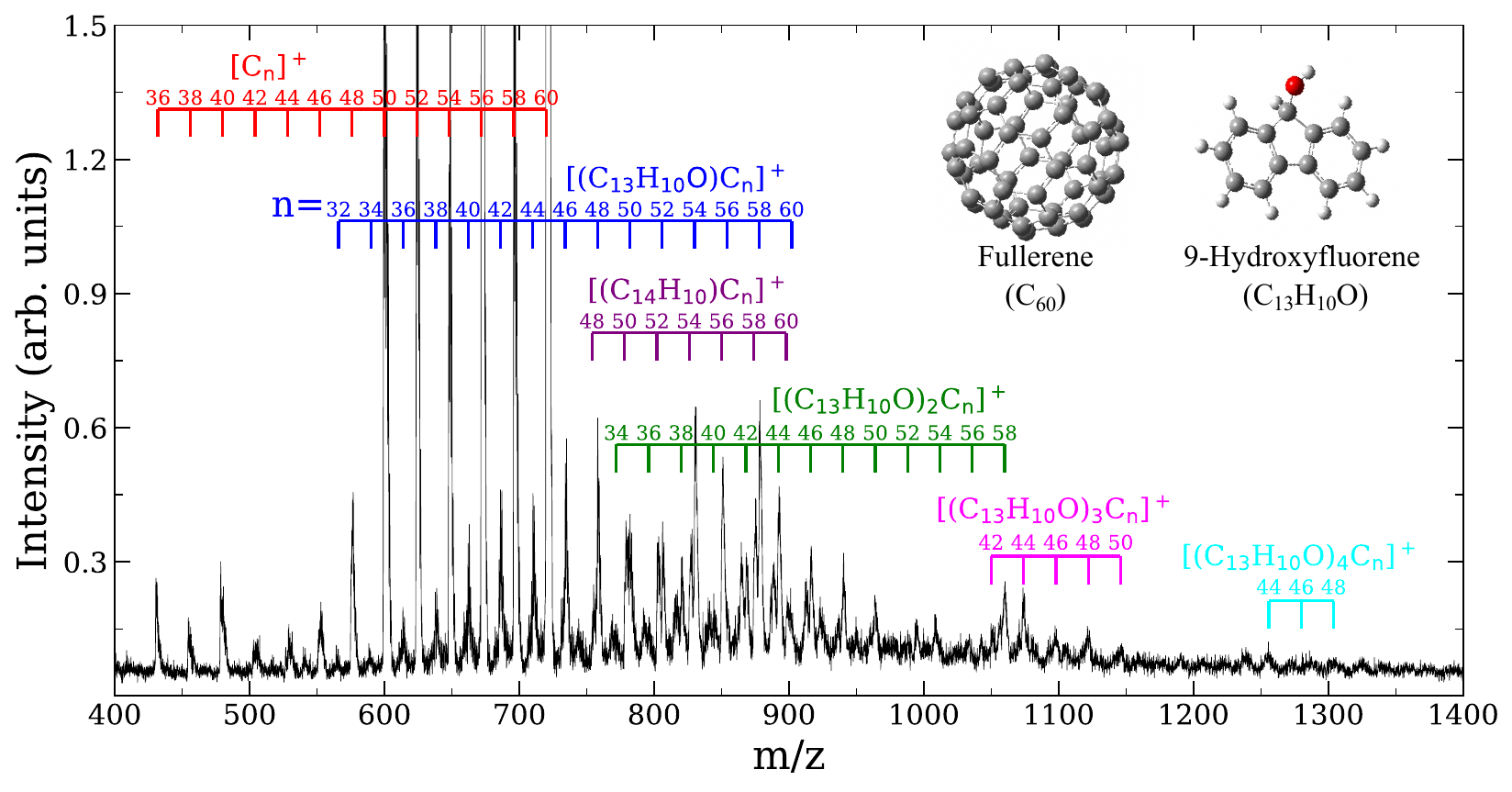}
	\vspace{-2.0em}
	\caption{The resulting mass spectrum of fullerene/9-hydroxyfluorene cluster cations. The structures of fullerene (C$_{60}$) and 9-hydroxyfluorene (C$_{13}$H$_{10}$O) are shown. The cation cluster series ([(C$_{13}$H$_{10}$O)$_{1-4}$C$_n$]$^+$) and their mass fragments are labeled, together with fullerene/anthracene cluster cations ([(C$_{14}$H$_{10}$)C$_n$]$^+$) resulting from contamination.
	}
	\vspace{-0.5em}
	\label{fig1}
\end{figure*}

In the interstellar medium (ISM) of galaxies, polycyclic aromatic hydrocarbon (PAH) molecules and their derivatives generally contribute to the infrared (IR) broadband features at 3.3, 6.2, 7.7, 8.6, and 11.2 $\mu$m \citep[e.g.][]{sel84,all89,pug89}. These molecules have been reported to be abundant and contain about 10\% of the elemental carbon in space, and they play a significant role in the ionization and energy balance of ISM \citep[and reference therein]{tie13}. Subsequently, they are further processed for hundreds of millions of years in the harsh environment of the ISM, in which various types of functional groups may be acquired, leading to functionalized PAHs, such as methyl (-CH$_3$), vinyl (-CHCH$_2$), methoxy (-OCH$_3$), amino (-NH$_2$), cyano and isocyano (-CN, -NC), acid (-COOH), and hydroxyl (-OH) \citep{hol99,ber02,all11}. Recently, two nitrile-group-functionalized PAHs, 1- and 2-cyano naphthalene, were detected in the ISM. Both bicyclic ring PAH molecules were observed and confirmed in the TMC-1 molecular cloud \citep{mcg21}. 

It has been reported that fullerenes, C$_{60}$ and C$_{70}$, also exist in space according to the IR spectra of circumstellar and interstellar sources \citep{cam10,sel10}. Extensive experimental and theoretical investigations reveal that fullerene molecules may be formed by large PAHs through photochemical evolution in interstellar environments \citep{ber12,zhen2014,omo16,can19}. Moreover, several near-IR diffuse interstellar bands (DIBs) were linked to the electronic transitions of fullerene cation C$_{60}$$^+$ \citep{cam15,wal15,cor17}, providing new insights into understanding the chemical complexity of ISM \citep{cam18,cor19}. The stability of fullerenes with large C-atom counts ($n$$=$44, 45, ..., 70) was theoretically investigated, which suggested that smaller fullerenes with C-atom counts of 56, 50, and 44 possibly exist in astrophysical sources \citep{can19}.

In interstellar space where PAH and fullerene molecules coexist, fullerene/PAH adducts and their associated derivate clusters may be formed through ion-molecule reactions in the gas phase, resulting in the existence of large fullerene/PAH derived clusters \citep{pet93,dun13,gar13b,boh16,omo16,zhen2019b}. Furthermore, the size of some large fullerene/PAH clusters can fall into the range of a few nanometers, and when they are condensed, they may be further incorporated into cosmic grains \citep{zhen2019b,hu21a}. Therefore, studies on the formation of fullerene/PAHs clusters can offer a possible formation pathway for the grains, and provide insight for understanding the evolution of carbon-rich molecules \citep{mar20}. 

The ionization or chemical states of interstellar PAHs, fullerenes, and their derivatives are expected to be significantly affected by their chemical-physical environments \citep{bak94,lep01}. To understand the chemical characteristics of interstellar fullerenes and PAHs, the differential chemical reactivity of fullerenes and various types of PAHs are needed to investigate, e.g., PAHs with substituted functional groups \citep{boh16,zhen2019a,zhen2019b,hu21a,hu21b}.

To understand the reaction between fullerene cations and oxygenated groups substituted PAH molecules, and further investigate the coevolution network of interstellar fullerene chemistry \citep{tie13,omo16}, in this work, we present an exploration study of the chemical reactivity of the fullerene cations (C$_n$$^+$, $n$$=$32, 34, ..., 60) with 9-hydroxyfluorene (C$_{13}$H$_{10}$O, 24 atoms, 182 amu). Furthermore, to investigate the formation mechanism of fullerene/9-hydroxyfluorene cluster cations, the experimental results are illustrated together with the results of the theoretical chemistry calculation. 

The molecule 9-hydroxyfluorene, the structure of which is shown in the top right corner of Fig~\ref{fig1}, is selected as a typical example of interstellar PAH molecules based on the following consideration: (1) 9-hydroxyfluorene has a relatively large size among the organic molecules contained with oxygenated functional groups; (2) 9-hydroxyfluorene is suitable for heating in the oven that can efficiently sublimates into the gas phase. 

\begin{figure*}
	\centering
	\includegraphics[width=4.8in]{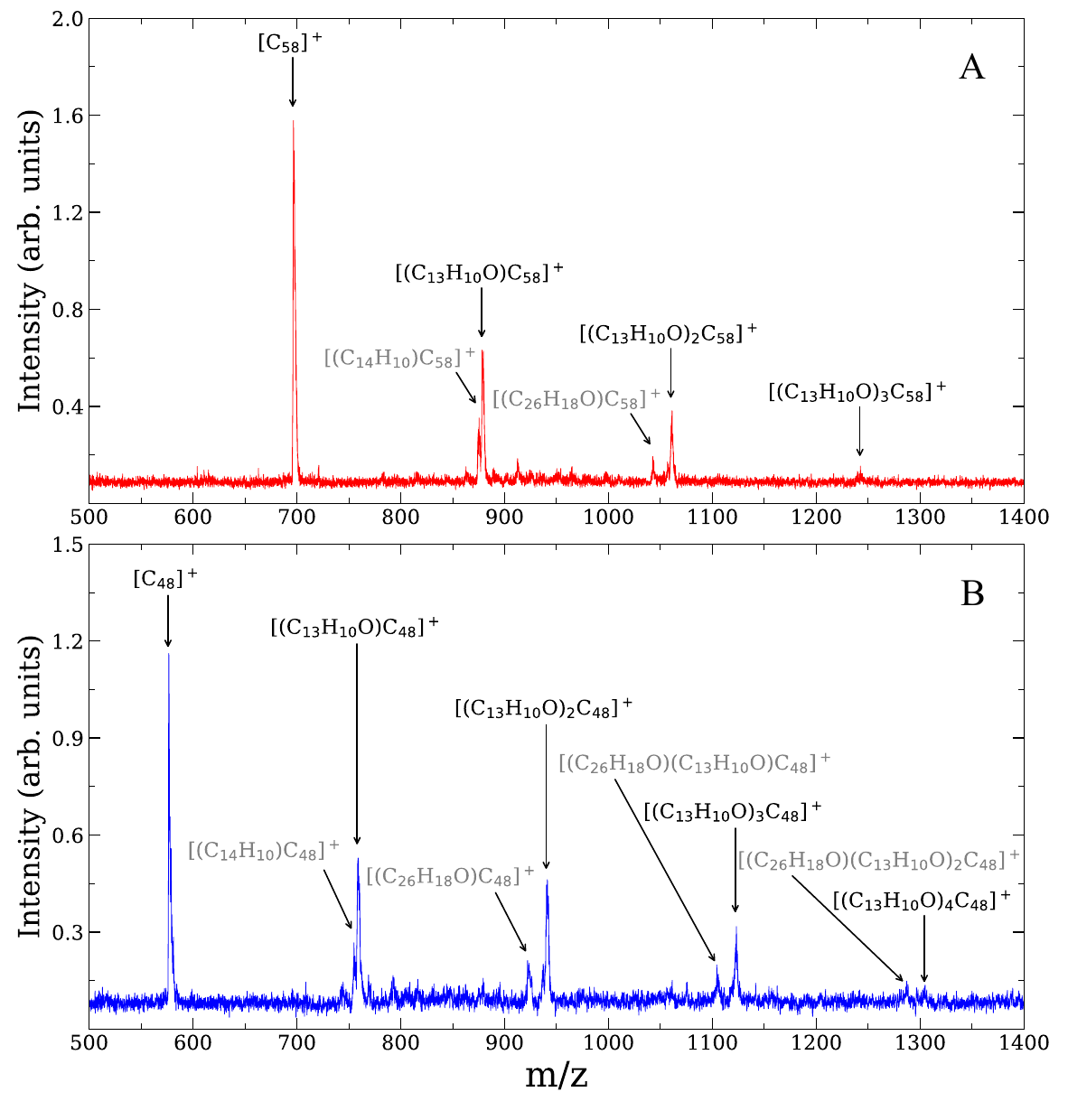}
	\vspace{-0.8em}
	\caption{The resulting mass spectra of C$_{58}$$^+$/9-hydroxyfluorene cluster cations (panel A), and C$_{48}$$^+$/9-hydroxyfluorene cluster cations (panel B) when using the SWIFT technique. The cluster cation series are labeled (black), together with the fullerene/dimer clusters and contamination of fullerene/anthracene cluster cations (gray). 
	}
	\label{fig2}
	\vspace{-0.5em}
\end{figure*}

\section{Experimental results}
\label{sub:expt}

The experiments were performed on an experimental apparatus equipped with a quadrupole ion trap and reflection time-of-flight (QIT-TOF) mass spectrometer \citep[details provided in Appendix A,][]{zhen2019a,zhen2019b}. C$_{60}$$^+$ was generated by electron impact ionization and filled into the trap within the time range of 0.0$-$4.0 s. Then a 355 nm laser ($\sim$ 20 mJ/pulse, 10 Hz, irradiation time amounting to 1.8 s, from 4.0$-$5.8 s) was used to irradiate the trapped C$_{60}$$^+$ to generate smaller fullerene cations. Under laser irradiation, the trapped fullerene cations underwent photo-fragmentation processes, i.e., they successively lost C$_2$ units to form a series of smaller fullerene cations \citep[C$_{n}$$^+$, $n$$=$32, 34, ..., 58,][]{lif00,zhen2014}. These newly formed fullerene cations subsequently reacted with gas-phase neutral 9-hydroxyfluorene molecules to form fullerene/9-hydroxyfluorene cluster cations (5.8$-$9.88 s), and then detected by the reflection time-of-flight mass spectrometer. All the species are observed as peaks with different mass/charge ($m/z$) in the mass spectrum, and can be identified given the initial reactants. 

Fig.~\ref{fig1} presents the resulting mass spectrum of fullerene/9-hydroxyfluorene cluster cations, showing that a series of cationic clusters have been generated. Fullerene cations (C$_{n}$$^+$, $n$$=$36, 38, ..., 60; marked in red), and four groups of fullerene/9-hydroxyfluorene cluster cations are presented and labeled: mono-C$_{13}$H$_{10}$O adducts ([(C$_{13}$H$_{10}$O)C$_n$]$^+$, $n$$=$32, 34, ..., 60; marked in blue), di-C$_{13}$H$_{10}$O adducts ([(C$_{13}$H$_{10}$O)$_2$C$_{n}$]$^+$, $n$$=$32, 34, ..., 58; marked in green), tri-C$_{13}$H$_{10}$O adducts ([(C$_{13}$H$_{10}$O)$_3$C$_{n}$]$^+$, $n$$=$42, 44, ..., 50; marked in magenta), and tetra-C$_{13}$H$_{10}$O adducts ([(C$_{13}$H$_{10}$O)$_4$C$_{n}$]$^+$, $n$$=$44, 46, 48; marked in cyan). Some mass peaks of species with similar $m/z$ range may overlap. We also observed and labeled some additional mass peaks, which were formed as a side-product due to contaminations of anthracene (C$_{14}$H$_{10}$) in the ion trap chamber \citep{zhen2019b}. For example, one series of additional peaks can be assigned as [(C$_{14}$H$_{10}$)C$_{n}$]$^+$ ($n$$=$48, 50, ..., 60; marked in purple). 

Based on the observed new species, we supposed that fullerene/9-hydroxyfluorene cluster cations are formed through collision reactions between fullerene cations (C$_{n}$$^+$, $n$$=$32, 34, ..., 60) and neutral C$_{13}$H$_{10}$O molecules, similarly to previous work on fullerene-PAH cationic clusters \citep{boh16,zhen2019b}. The reaction process of fullerene cations and C$_{13}$H$_{10}$O was carried out through sequential steps, in which C$_{13}$H$_{10}$O molecules were added repeatedly to the surface of fullerene cation cages \citep{gar13b,sat13,zhen2019a}. As a result, a series of large fullerene/9-hydroxyfluorene cluster cations were formed. We note that only one C$_{13}$H$_{10}$O molecule can be added to C$_{60}$$^+$, while with the size of fullerene species decreased, more C$_{13}$H$_{10}$O molecules can be added to smaller fullerene cations, suggesting an enhanced chemical reactivity for smaller fullerene cations. 

The corresponding formation pathways for fullerene/9-hydroxyfluorene cluster cations are summarized below (equations 1-4):

\vspace{-1.5em}
\begin{equation}
	[{\rm C}_{n}]^+ \xrightarrow{{\rm C}_{13}{\rm H}_{10}{\rm O}} [({\rm C}_{13}{\rm H}_{10}{\rm O}){\rm C}_n]^+
\end{equation}

\vspace{-2.0em}
\begin{equation}
	[({\rm C}_{13}{\rm H}_{10}{\rm O}){\rm C}_n]^+ \xrightarrow{{\rm C}_{13}{\rm H}_{10}{\rm O}} [({\rm C}_{13}{\rm H}_{10}{\rm O})_2{\rm C}_n]^+
\end{equation}

\vspace{-2.0em}
\begin{equation}
	[({\rm C}_{13}{\rm H}_{10}{\rm O})_2{\rm C}_n]^+ \xrightarrow{{\rm C}_{13}{\rm H}_{10}{\rm O}} [({\rm C}_{13}{\rm H}_{10}{\rm O})_3{\rm C}_n]^+
\end{equation}

\vspace{-2.0em}
\begin{equation}
	[({\rm C}_{13}{\rm H}_{10}{\rm O})_3{\rm C}_n]^+ \xrightarrow{{\rm C}_{13}{\rm H}_{10}{\rm O}} [({\rm C}_{13}{\rm H}_{10}{\rm O})_4{\rm C}_n]^+
\end{equation}

Due to the weak intensity and overlap with each other, some mass peaks cannot be identified in Fig.~\ref{fig1}. To clarify that, we performed another experiment adopting the stored waveform inverse Fourier transform excitation (SWIFT) isolation technique \citep{dor96}. In this experiment, when the laser irradiation ended, a SWIFT pulse was applied to the end caps of the ion trap, allowing for the selection of the desired fullerene cations with a specific mass/charge ($m/z$) range to understand their reactions.

Fig.~\ref{fig2} displays two typical mass spectra for the clusters formed through reactions between individual fullerene cations (here, C$_{58}$$^+$ and C$_{48}$$^+$) and 9-hydroxyfluorene molecules, where the newly formed clusters for C$_{58}$$^+$ and C$_{48}$$^+$ were observed and labeled in Fig.~\ref{fig2}(A) and Fig.~\ref{fig2}(B), respectively. The observed mass peaks for fullerene cations (C$_{58}$$^+$/C$_{48}$$^+$), fullerene/9-hydroxyfluorene clusters ([(C$_{13}$H$_{10}$O)$_{1-3}$C$_{58}$]$^+$ and [(C$_{13}$H$_{10}$O)$_{1-4}$C$_{48}$]$^+$) are identified, together with fullerene/anthracene cluster cations ([(C$_{14}$H$_{10}$)C$_{58}$]$^+$ and [(C$_{14}$H$_{10}$)C$_{48}$]$^+$) resulting from contamination. The mass spectra when using the SWIFT technique for the two individual fullerene cations show a similar reaction process, implying the sequential steps for the reactions of fullerene cations with C$_{13}$H$_{10}$O molecules. 

Interestingly, we also observed several additional peaks in Fig.~\ref{fig2}. These additional peaks are formed as side-products due to the neutral dimer 9-hydroxyfluorene molecule (C$_{26}$H$_{18}$O, $m$/$z$=346) that formed in the heat processes trough dehydration pathways. For example, as shown in Fig.~\ref{fig2}(B), three additional peaks can be assigned as: [(C$_{26}$H$_{18}$O)C$_{48}$]$^+$ ($m/z$=922), [(C$_{26}$H$_{18}$O)(C$_{13}$H$_{10}$O)C$_{48}$]$^+$ ($m/z$=1104), and [(C$_{26}$H$_{18}$O)(C$_{13}$H$_{10}$O)$_2$C$_{48}$]$^+$ ($m/z$=1286). The corresponding formation pathways for the fullerene/dimer cluster cations are summarized below (equations 5-6):

\vspace{-1.0em}
\begin{equation}
	[{\rm C}_{n}]^+ \xrightarrow{{\rm C}_{26}{\rm H}_{18}{\rm O}} [({\rm C}_{26}{\rm H}_{18}{\rm O}){\rm C}_n]^+
\end{equation}

\vspace{-2.0em}
\begin{equation}
	[({\rm C}_{26}{\rm H}_{18}{\rm O}){\rm C}_n]^+ \xrightarrow{{\rm C}_{13}{\rm H}_{10}{\rm O}} [({\rm C}_{26}{\rm H}_{18}{\rm O})({\rm C}_{13}{\rm H}_{10}{\rm O}){\rm C}_n]^+ 
\end{equation}

\section{Theoretical calculation results}
\label{sub:theo}

To understand the formation mechanism of the fullerene/9-hydroxyfluorene cluster cation system, the reaction channels occurring on the fullerene surface during the accretion processes were theoretically studied. Here, the structures of fullerene (C$_{60}$) cation and a defective fullerene (C$_{58}$) cation along with C$_{13}$H$_{10}$O and its dimer formed by dehydration reactions (C$_{26}$H$_{18}$O), and the reaction pathways of C$_{60}$$^+$ $+$ C$_{13}$H$_{10}$O and C$_{58}$$^+$ $+$ C$_{13}$H$_{10}$O/C$_{26}$H$_{18}$O were performed through quantum theoretical calculations. 

All the calculations were performed by the Gaussian 16 program \citep{fri16} package, using density functional theory employing the hybrid functional B3LYP \citep{bec92,lee88} with the 6-311++G(d,p) basis set, which is consistent with previous study \citep[e.g.,][]{hu23}. We followed the minimum energy pathway of the Van der Waals cluster or the covalently bonded cluster and, at each step, calculated the molecular geometries and binding energies. The negative and positive binding energies correspond to exothermic and endothermic reactions. Since we use the helium atoms as a stabilization \citep{zhen2019a,zhen2019b}, the molecular collision in our laboratory is at a low velocity. Thus, we think the exothermic reactions are dominant and easier to occur, while the endothermic reactions could not occur. To account for the intermolecular interactions, dispersion correction \citep[D3,][]{gri11} was considered for each system. The vibrational frequencies were calculated for the optimized geometries to verify that these geometries correspond to the minima on the potential energy surface. Furthermore, the zero-point vibrational energy was obtained from the frequency calculation to correct the molecular energy. 

For fullerene cations (C$_{60}$$^+$ and C$_{58}$$^+$), the molecular structures were mainly consistent with that in our previous work \citep{zhen2019a,zhen2019b,hu21a,hu23}. In this work, we assumed that there was only the C$_2$ loss at a local position and no carbon skeleton rearrangement during the electron impact ionization and fragmentation \citep{zhen2019b,can19}. Fullerene cations (C$_{60}$$^+$ and C$_{58}$$^+$) and their derivate cationic clusters have an odd number of electrons, resulting in an open-shell doublet ground state (spin multiplicity of 2) for the ground electronic state. The neutral C$_{13}$H$_{10}$O and C$_{26}$H$_{18}$O have closed-shell electronic structures, and their spin multiplicity is 1. 

\begin{figure}
	\centering
	\includegraphics[width=3.6in]{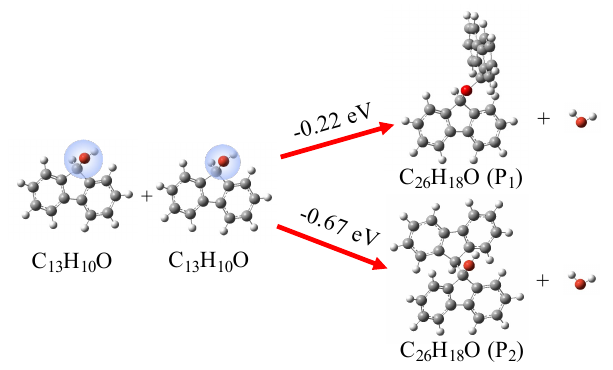}
	\vspace*{-2.0em}
	\caption{The formation pathways and optimized geometries of C$_{26}$H$_{18}$O. The negative energy is corresponding to the exothermic reaction.
	}
	\label{fig3}
	\vspace*{-1.8em}
\end{figure}

The calculated results for the formation pathways and optimized geometries of C$_{26}$H$_{18}$O are shown in Fig.~\ref{fig3}. The calculation results for C$_{60}$$^+$ $+$ C$_{13}$H$_{10}$O are shown in Fig.~\ref{fig4}. The calculation results for C$_{58}$$^+$ with C$_{13}$H$_{10}$O are provided in Fig.~\ref{fig5}-\ref{fig8}. The calculation results for C$_{58}$$^+$ $+$ C$_{26}$H$_{18}$O are provided in Fig.~\ref{fig9}.

\subsection{The formation pathways and optimized geometries of C$_{26}$H$_{18}$O}

The formation pathways and optimized geometries of C$_{26}$H$_{18}$O are illustrated in Fig.~\ref{fig3}. C$_{26}$H$_{18}$O is mainly formed through an intermolecular dehydration pathway during heating in the oven. To form C$_{26}$H$_{18}$O (P$_{1}$), an ether (C-O-C) bond is newly formed with exothermic energy of $-$0.22 eV ($-$5.0 kcal mol$^{-1}$); for the formation of C$_{26}$H$_{18}$O (P$_{2}$), a C-C single bond is newly formed with exothermic energy of $-$0.67 eV ($-$15.5 kcal mol$^{-1}$). Based on the calculation results, both molecules C$_{26}$H$_{18}$O (P$_{1}$ \& P$_{2}$) are selected as the dehydration products of C$_{13}$H$_{10}$O for the further calculations of the cluster formation.

\subsection{Calculated results of C$_{60}$$^+$ $+$ C$_{13}$H$_{10}$O}

\begin{figure}
	\centering
	\includegraphics[width=2.8in]{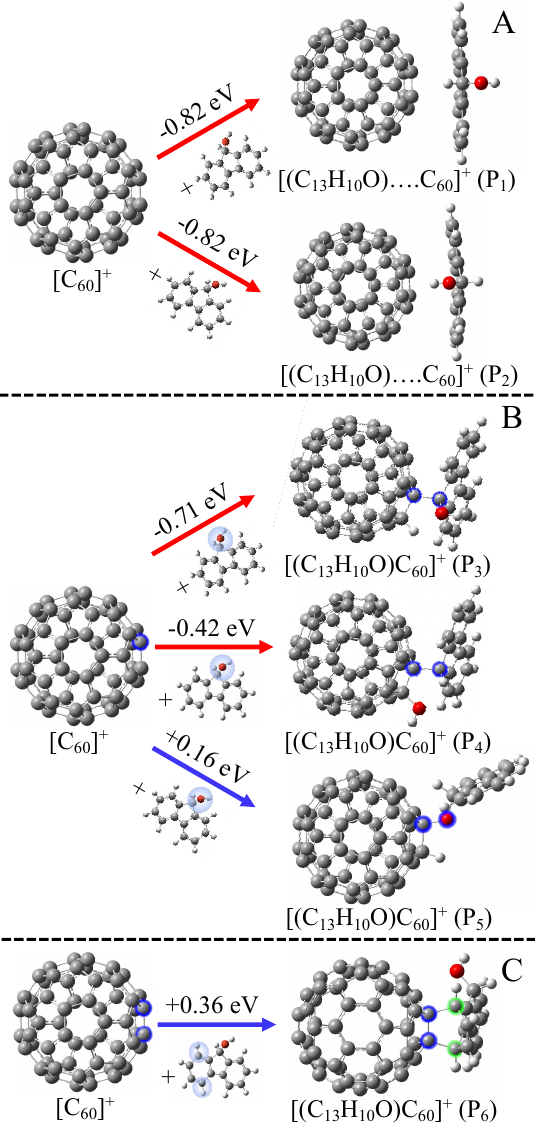}
	\caption{The reaction pathways of C$_{60}$$^+$ $+$ C$_{13}$H$_{10}$O. The negative and positive binding energies correspond to exothermic and endothermic reactions.
	}
	\vspace*{-0.8em}
	\label{fig4}
\end{figure}

Fig.~\ref{fig4} depicts the formation channel of the adduct between C$_{60}$$^+$ and C$_{13}$H$_{10}$O. Fig.~\ref{fig4}(A) is for the formation of the Van der Waals clusters, and Fig.~\ref{fig4}(B \& C) for the formation of the covalently bonded clusters, consistent with previous works \citep[e.g.][]{zhen2019b,hu21b,hu23}

As shown in Fig.~\ref{fig4}(A), the Van der Waals cluster molecules, [(C$_{13}$H$_{10}$O)....C$_{60}$]$^+$ (P$_1$) and [(C$_{13}$H$_{10}$O)....C$_{60}$]$^+$ (P$_2$) are formed in two different molecular alignments. The reactions are exothermic by the same energy of $-$0.82 eV ($-$18.9 kcal mol$^{-1}$). Therefore, the molecular alignments, e.g., H-C-OH or HO-C-H, do not affect the formation of the Van der Waals cluster molecules. Furthermore, in order to better demonstrate the binding characteristics of these newly formed molecular clusters, the three-dimensional (3D) images of the Mulliken charge separation, highest occupied molecular orbital (HOMO) and lowest unoccupied molecular orbital (LUMO) of these molecules are obtained and provided in the Appendix B (Fig.~\ref{figB1}), respectively.

In Fig.~\ref{fig4}(B), three possible reaction pathways concerning the functional groups, -CHOH unit, of C$_{13}$H$_{10}$O are obtained. For the formation of [(C$_{13}$H$_{10}$O)C$_{60}$]$^+$ (P$_3$), C$_{60}$$^+$ and C$_{13}$H$_{10}$O are connected by a newly formed C-C single bond, in which one C atom is from C$_{60}$$^+$ and the other from -CHOH group. Meanwhile, the H atom from the CH unit in the -CHOH group directly migrates to the nearby carbon on the surface of C$_{60}$$^+$, forming a new C-H bond. The reaction has an exothermic reaction energy of $-$0.71 eV ($-$16.4 kcal mol$^{-1}$). To form [(C$_{13}$H$_{10}$O)C$_{60}$]$^+$ (P$_4$), C$_{60}$$^+$ and C$_{13}$H$_{10}$O are connected by a newly formed C-C single bond, in which one C atom is from C$_{60}$$^+$ and the other from -CHOH group. Meanwhile, the OH unit in the -CHOH group directly migrates to the nearby carbon on the surface of C$_{60}$$^+$, forming a new C-OH bond. The reaction is exothermic by $-$0.42 eV ($-$9.7 kcal mol$^{-1}$). To form [(C$_{13}$H$_{10}$O)C$_{60}$]$^+$ (P$_5$), C$_{60}$$^+$ and C$_{13}$H$_{10}$O are connected by a newly formed C-O single bond, in which the C atom is from C$_{60}$$^+$ and the O atom from -CHOH group. Meanwhile, the H atom from the OH group in the -CHOH group directly migrates to the nearby carbon on the surface of C$_{60}$$^+$, forming a new C-H bond. The reaction is endothermic by $+$0.16 eV ($+$3.7 kcal mol$^{-1}$). Thus, we suppose this reaction pathway cannot occur in the ISM. 

Moreover, concerning the aromatic ring of C$_{13}$H$_{10}$O, the possible reaction pathway of [(C$_{13}$H$_{10}$O)C$_{60}$]$^+$ (P$_6$) is illustrated in Fig.~\ref{fig4}(C), in which C$_{60}$$^+$ and C$_{13}$H$_{10}$O are connected by double C-C bridge bonds, similar to previous works \citep{pet93,sat13,zhen2019b}. The reaction is endothermic by $+$0.36 eV ($+$8.3 kcal mol$^{-1}$), and we suppose it is unlikely to occur.

The exothermic energy for the formation of [(C$_{13}$H$_{10}$O)C$_{60}$]$^+$ (P$_1$-P$_4$), with binding energies between $-$0.42 and $-$0.82 eV ($-$9.7 and $-$18.9 kcal mol$^{-1}$), is relatively higher and can stabilize the whole clusters. Thus, we propose that [(C$_{13}$H$_{10}$O)C$_{60}$]$^+$ formed in the laboratory is a mixed cluster with different possible isomers.

\subsection{Calculated results of C$_{58}$$^+$ $+$ C$_{13}$H$_{10}$O}

For the interaction of C$_{58}$$^+$ with C$_{13}$H$_{10}$O, due to the structure of C$_{58}$$^+$ \citep[e.g.,][]{lee04}, as shown in Fig.~\ref{fig6}, there are two types of reaction pathways needed to be considered: C$_{13}$H$_{10}$O ``landing'' on the 6 C-ring, and C$_{13}$H$_{10}$O ``landing'' on the 7 C-ring, which is in agreement with previous works \citep{zhen2019b,hu23}. The calculation results are presented in Fig.~\ref{fig5}-\ref{fig7}, respectively. 

\subsubsection{Calculated results of C$_{58}$$^+$ (6 C-ring) $+$ C$_{13}$H$_{10}$O}

\begin{figure}
	\centering
	\includegraphics[width=3.3in]{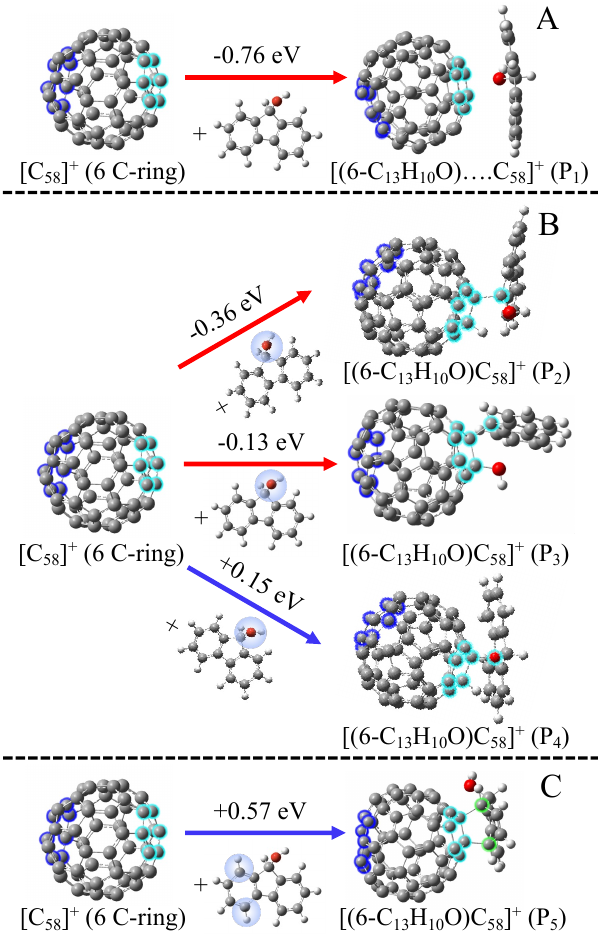}
	\caption{The reaction pathways of C$_{58}$$^+$ (6 C-ring) $+$ C$_{13}$H$_{10}$O. The negative and positive binding energies correspond to exothermic and endothermic reactions. The 7 C-ring and 6 C-ring are highlighted.
	}
	\label{fig5}
	\vspace*{-0.5em}
\end{figure}

For the reaction type of C$_{13}$H$_{10}$O ``landing'' on the 6 C-ring of C$_{58}$$^+$, similar to the reaction of C$_{60}$$^+$ $+$ C$_{13}$H$_{10}$O, the formation channel of [(6-C$_{13}$H$_{10}$O)C$_{58}$]$^+$ is demonstrated in Fig.~\ref{fig5}.

In Fig.~\ref{fig5}(A), the Van der Waals cluster molecules, [(6-C$_{13}$H$_{10}$O)....C$_{58}$]$^+$ (P$_1$) is formed in a typical molecular alignment, and the reaction is exothermic by $-$0.76 eV ($-$17.5 kcal mol$^{-1}$). Three possible reaction pathways concerning the -CHOH unit of C$_{13}$H$_{10}$O are obtained in Fig.~\ref{fig5}(B). [(6-C$_{13}$H$_{10}$O)C$_{58}$]$^+$ (P$_2$) is formed by the H transfer reaction with exothermic energy of $-$0.36 eV ($-$8.3 kcal mol$^{-1}$), in which C$_{58}$$^+$ and C$_{13}$H$_{10}$O are connected by a newly formed C-C single bond with H from CH group migrating to the nearby carbon on the surface of C$_{58}$$^+$ and forming a new C-H bond. To form [(6-C$_{13}$H$_{10}$O)C$_{58}$]$^+$ (P$_3$), C$_{58}$$^+$ and C$_{13}$H$_{10}$O are connected by a newly formed C-C single bond with OH migrating to the nearby carbon on the surface of C$_{58}$$^+$ and forming a new C-OH bond. The reaction is exothermic by $-$0.13 eV ($-$3.0 kcal mol$^{-1}$). [(6-C$_{13}$H$_{10}$O)C$_{58}$]$^+$ (P$_4$) is also formed by the H transfer reaction, in which the H atom is from the OH group. The reaction has an endothermic energy of $+$0.15 eV ($+$3.5 kcal mol$^{-1}$) and cannot occur. Moreover, [(6-C$_{13}$H$_{10}$O)C$_{58}$]$^+$ (P$_5$) illustrated in Fig.~\ref{fig5}(C) is the product concerning the aromatic ring of C$_{13}$H$_{10}$O, in which C$_{58}$$^+$ and C$_{13}$H$_{10}$O are connected by double C-C bonds. The reaction is endothermic by $+$0.57 eV ($+$13.1 kcal mol$^{-1}$) and cannot occur.

By comparing the formation of [(6-C$_{13}$H$_{10}$O)C$_{58}$]$^+$ with that of [(C$_{13}$H$_{10}$O)C$_{60}$]$^+$, we find no variation of the binding energy with the same reaction modes, implying no deformation for the 6 C-ring of C$_{58}$$^+$.

\subsubsection{Calculated results of C$_{58}$$^+$ (7 C-ring) $+$ C$_{13}$H$_{10}$O (I)}

\begin{figure*}
	\centering		
	\vspace*{-1.0em}
	\includegraphics[width=6.5in]{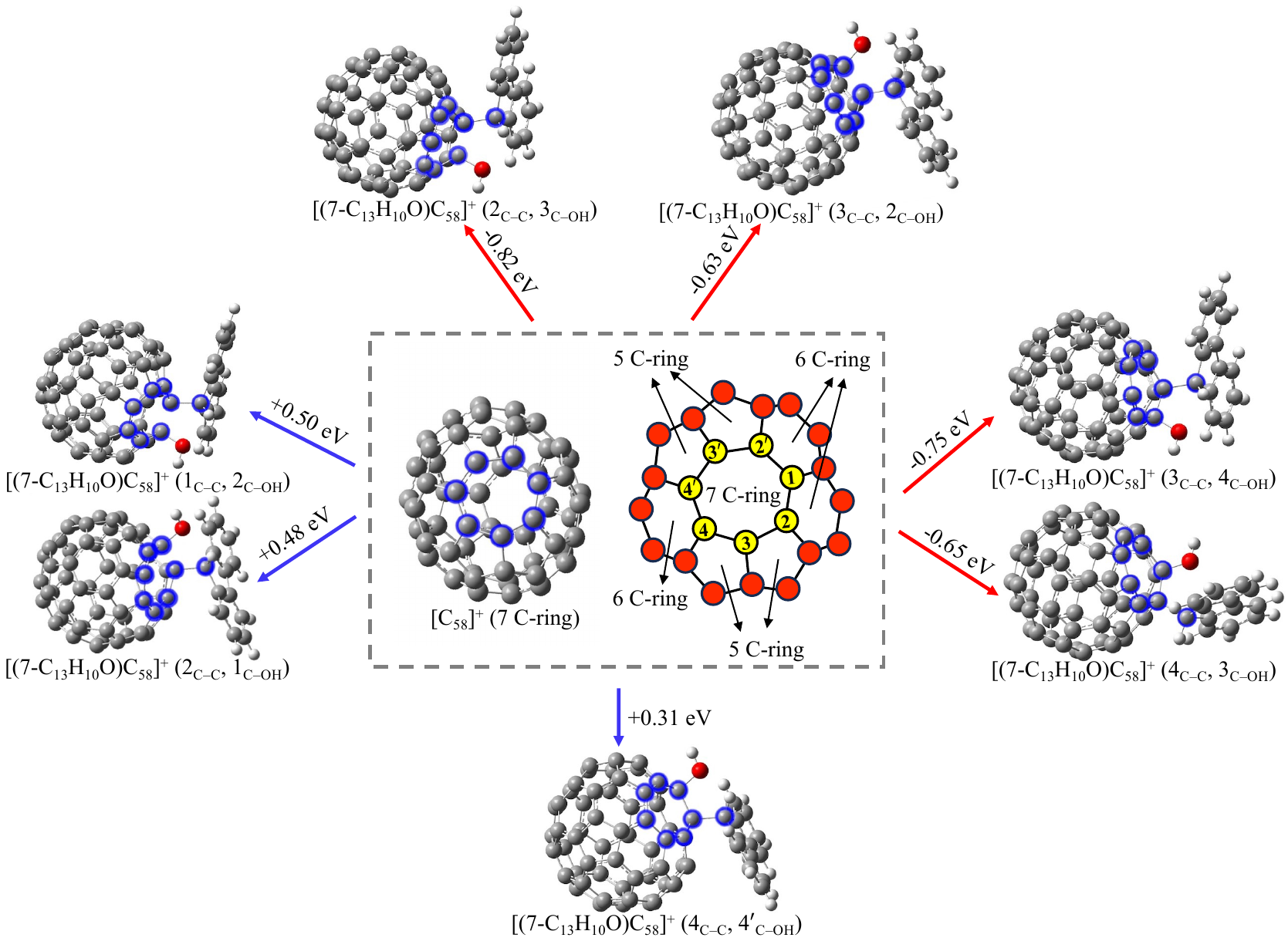}
	\caption{The OH transfer reactions between C$_{58}$$^+$ (7 C-ring) and C$_{13}$H$_{10}$O with different combination sites. The negative and positive binding energies correspond to exothermic and endothermic reactions. The 7 C-ring is highlighted.
	}
	\label{fig6}
	\vspace*{-1.0em}
\end{figure*}

Due to the multi-carbon reaction sites of C$_{58}$$^+$ (7 C-ring) that can be applied for the further adduct reactions, we carried out the chemical reactivity for these multi-carbon reaction sites of C$_{58}$$^+$ (7 C-ring) with C$_{13}$H$_{10}$O. Three factors are considered for the chemical reactivity, including the reaction mode with C$_{13}$H$_{10}$O, the combining carbon reaction site, and the reaction boundary. The obtained reaction pathways and optimized geometries of products are shown in Fig.~\ref{fig6}.

One typical reaction mode is selected: the OH group transfer reactions. Thus, these seven carbon combining reaction sites of C$_{58}$$^+$ (7 C-ring) can be divided and labeled as 1, 2 and 2$'$, 3 and 3$'$, 4 and 4$'$. Since the OH group transfer reaction pathways need two adjacent carbon sites along the boundary of the 7 C-ring, four combination modes can be obtained as (1,2)=(1,2$'$), (2,3)=(2$'$,3$'$), (3,4)=(3$'$,4$'$), (4,4$'$). (1,2)=(1,2$'$) and (4,4$'$) are the boundaries of the 7 C-ring and 6 C-ring, while (2,3)=(2$'$,3$'$) and (3,4)=(3$'$,4$'$) are the boundaries of the 7 C-ring and 5 C-ring.

As demonstrated in Fig.~\ref{fig6}, the combination modes of (2,3) and (3,4) have a higher reactivity than (1,2) and (4,4$'$). The formation reaction pathways of [(7-C$_{13}$H$_{10}$O)C$_{58}$]$^+$ (2$_{\rm C-C}$, 3$_{\rm C-OH}$ \& 3$_{\rm C-C}$, 2$_{\rm C-OH}$) or [(7-C$_{13}$H$_{10}$O)C$_{58}$]$^+$ (3$_{\rm C-C}$, 4$_{\rm C-OH}$ \& 4$_{\rm C-C}$, 3$_{\rm C-OH}$) are exothermic by $-$0.82 and $-$0.63 eV ($-$18.9 and $-$14.5 kcal mol$^{-1}$), or $-$0.75 and $-$0.65 eV ($-$17.3 and  $-$15.0 kcal mol$^{-1}$), respectively. However, the formation reaction pathways of [(7-C$_{13}$H$_{10}$O)C$_{58}$]$^+$ (4$_{\rm C-C}$,4$'$$_{\rm C-OH}$) or [(7-C$_{13}$H$_{10}$O)C$_{58}$]$^+$ (2$_{\rm C-C}$, 1$_{\rm C-OH}$ \& 1$_{\rm C-C}$, 2$_{\rm C-OH}$) are endothermic $+$0.31 eV ($+$7.2 kcal mol$^{-1}$) or $+$0.48 and $+$0.50 eV ($+$11.1 and $+$11.5 kcal mol$^{-1}$), respectively, and unlikely to occur. 

Accordingly, the reaction tends to occur along the boundary of the 7 C-ring and 5 C-ring rather than the boundary of the 7 C-ring and 6 C-ring, suggesting that the combination mode of (2,3)=(3,4) has a higher chemical reactivity, which is in agreement with \citet{pet93}. Therefore, we choose the combination mode of (2,3) for further calculations of the cluster formation. 

\subsubsection{Calculated results of C$_{58}$$^+$ (7 C-ring) $+$ C$_{13}$H$_{10}$O (II)}

\begin{figure}
	\centering
	\includegraphics[width=3.0in]{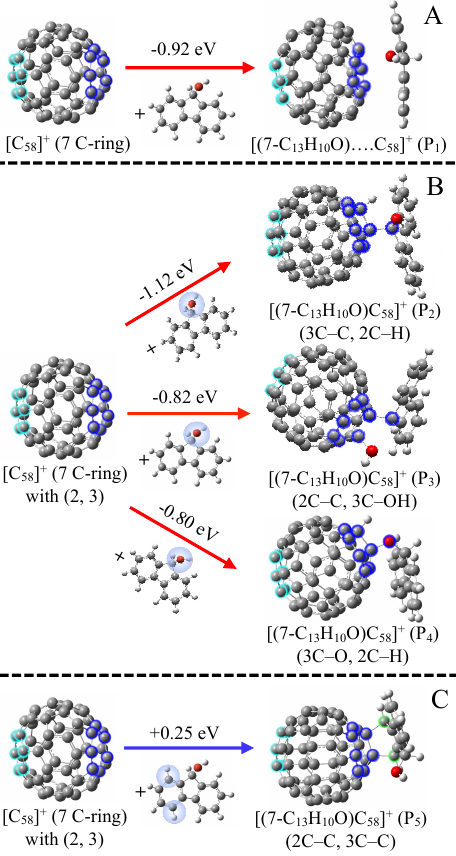}
	\vspace*{-0.5em}
	\caption{The reaction pathways of C$_{58}$$^+$ (7 C-ring) with (2,3) $+$ C$_{13}$H$_{10}$O. The negative and positive binding energies correspond to exothermic and endothermic reactions. The 7 C-ring and 6 C-ring are highlighted.
	}
	\label{fig7}
	\vspace*{-0.8em}
\end{figure}

Based on the calculated results, the reaction pathways of C$_{58}$$^+$ (7 C-ring) with (2,3) $+$ C$_{13}$H$_{10}$O for different reaction modes are obtained and illustrated in Fig.~\ref{fig7}. 

In Fig.~\ref{fig7}(A), the Van der Waals cluster molecule, [(7-C$_{13}$H$_{10}$O)....C$_{58}$]$^+$ (P$_1$) is formed in one typical molecular alignment, and the reaction is exothermic by $-$0.92 eV ($-$21.2 kcal mol$^{-1}$). In Fig.~\ref{fig7}(B), three possible reaction pathways concerning the -CHOH unit of C$_{13}$H$_{10}$O are obtained. To form [(7-C$_{13}$H$_{10}$O)C$_{58}$]$^+$ (P$_2$), C$_{58}$$^+$ and C$_{13}$H$_{10}$O are connected by a newly formed C-C single bond with the transfer of H from CH group to the adjacent carbon on the 7 C-ring surfaces of C$_{58}$$^+$ and the formation of a C-H single bond. The reaction has an exothermic energy of $-$1.12 eV ($-$25.8 kcal mol$^{-1}$). The formation pathways of [(7-C$_{13}$H$_{10}$O)C$_{58}$]$^+$ (P$_3$ \& P$_4$), with the transfer of OH unit and H atom from OH group, have an exothermic reaction energy of $-$0.82 and $-$0.80 eV ($-$18.9 and $-$18.4 kcal mol$^{-1}$), respectively. Moreover, [(7-C$_{13}$H$_{10}$O)C$_{58}$ (P$_5$) illustrated in Fig.~\ref{fig7}(C) is the product concerning the aromatic ring of C$_{13}$H$_{10}$O, in which C$_{58}$$^+$ and C$_{13}$H$_{10}$O are connected by double C-C bonds. The reaction is endothermic by $+$0.25 eV ($+$5.8 kcal mol$^{-1}$) and cannot occur. 

The exothermic energy of [(7-C$_{13}$H$_{10}$O)C$_{58}$]$^+$ is much higher than that of [(C$_{13}$H$_{10}$O)C$_{60}$]$^+$ or [(6-C$_{13}$H$_{10}$O)C$_{58}$]$^+$ with the same reaction modes, suggesting that there is an enhanced chemical reactivity of the 7 C-ring in C$_{58}$$^+$, consisting well with previous works \citep[e.g.][]{zhen2019b,hu21b}. Both the Van der Waals clusters, [(7-C$_{13}$H$_{10}$O)....C$_{58}$]$^+$ (P$_1$) and the covalently bonded clusters, [(7-C$_{13}$H$_{10}$O)C$_{58}$]$^+$ (P$_2$-P$_4$) have relatively high exothermic energy and may exist. 

Combining the calculation results for [(6-C$_{13}$H$_{10}$O)C$_{58}$]$^+$, we propose that [(C$_{13}$H$_{10}$O)C$_{58}$]$^+$ formed in the laboratory is also a mixed cluster with different possible isomers. Moreover, C$_{58}$$^+$ has a higher chemical reactivity than C$_{60}$$^+$, which is mainly attributed to the newly formed deformed 7 C-ring. 

\subsection{Calculated results of [(C$_{13}$H$_{10}$O)C$_{58}$]$^+$ $+$ C$_{13}$H$_{10}$O}{}

\begin{figure*}
	\centering
	\includegraphics[width=6.6in]{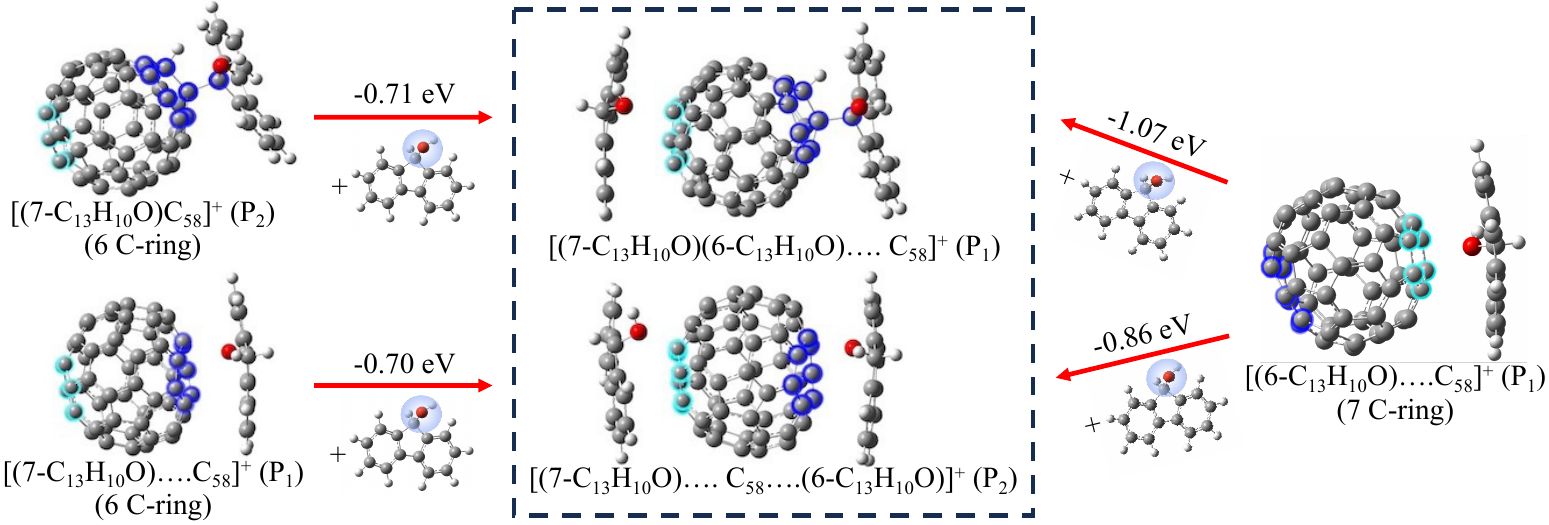}
	\caption{The reaction pathways of di-C$_{13}$H$_{10}$O adducts, (C$_{13}$H$_{10}$O)$_{2}$C$_{58}$$^+$. Both [(7-C$_{13}$H$_{10}$O)C$_{58}$]$^+$ (6 C-ring) $+$ C$_{13}$H$_{10}$O and [(6-C$_{13}$H$_{10}$O)C$_{58}$]$^+$ (7 C-ring) $+$ C$_{13}$H$_{10}$O were taken into consideration. The negative binding energy corresponds to the exothermic reaction.  The 7 C-ring and 6 C-ring are highlighted.
	}
	\label{fig8}
\end{figure*}

The formation routes for larger fullerene/9-hydroxyfluorene cluster cations were also investigated to elucidate the formation mechanism for larger C$_{58}$$^+$/9-hydroxyfluorene clusters. Fig.~\ref{fig8} displays the reaction routes for the formation of di-C$_{13}$H$_{10}$O adducts, (C$_{13}$H$_{10}$O)$_{2}$C$_{58}$$^+$, with considered three isomers of (C$_{13}$H$_{10}$O)C$_{58}$$^+$.

Due to the spatial effect, we suppose that the second C$_{13}$H$_{10}$O will be added on the 6 C-ring surfaces opposite to the 7 C-ring of [(7-C$_{13}$H$_{10}$O)C$_{58}$]$^+$ through the Van der Waals bond. {[}(7-C$_{13}$H$_{10}$O)(6-C$_{13}$H$_{10}$O)....C$_{58}${]}$^+$ (P$_1$) and {[}(7-C$_{13}$H$_{10}$O)....C$_{58}$....(6-C$_{13}$H$_{10}$O){]}$^+$ (P$_2$) are the products formed by the reaction pathways described above. The reactions are exothermic by $-$0.71 and $-$0.70 eV ($-$16.4 and $-$16.1 kcal mol$^{-1}$), respectively. We note that on the 6 C-ring surfaces, there is no variation in binding energies formed by C$_{58}$$^+$ or [(7-C$_{13}$H$_{10}$O)C$_{58}${]}$^+$ with C$_{13}$H$_{10}$O. 

Similarly, we obtained the reaction pathways from [(6-C$_{13}$H$_{10}$O)....C$_{58}$]$^+$ (P$_1$) to {[}(7-C$_{13}$H$_{10}$O)(6-C$_{13}$H$_{10}$O)....C$_{58}${]}$^+$ (P$_1$) and {[}(7-C$_{13}$H$_{10}$O)....C$_{58}$....(6-C$_{13}$H$_{10}$O){]}$^+$ (P$_2$), with high exothermic energy of $-$1.07 and $-$0.86 eV ($-$24.7 and $-$19.8 kcal mol$^{-1}$), respectively, suggesting the diversity of the reaction pathways for the formation of larger fullerene/9-hydroxyfluorene cluster cations.

The exothermic energy for the formation of [(C$_{13}$H$_{10}$O)$_{2}$C$_{58}$]$^+$ is higher, without a difference when compared with the formation of [(C$_{13}$H$_{10}$O)C$_{58}$]$^+$. Thus, consistent with the experimental results, (C$_{13}$H$_{10}$O)$_{2}$C$_{58}$$^+$ can easily formed in the laboratory and is a mixed cluster with different possible isomers.

\subsection{Calculated results of C$_{58}$$^+$ (7 C-ring) $+$ C$_{26}$H$_{18}$O}

\begin{figure*}
	\centering
	\includegraphics[width=6.6in]{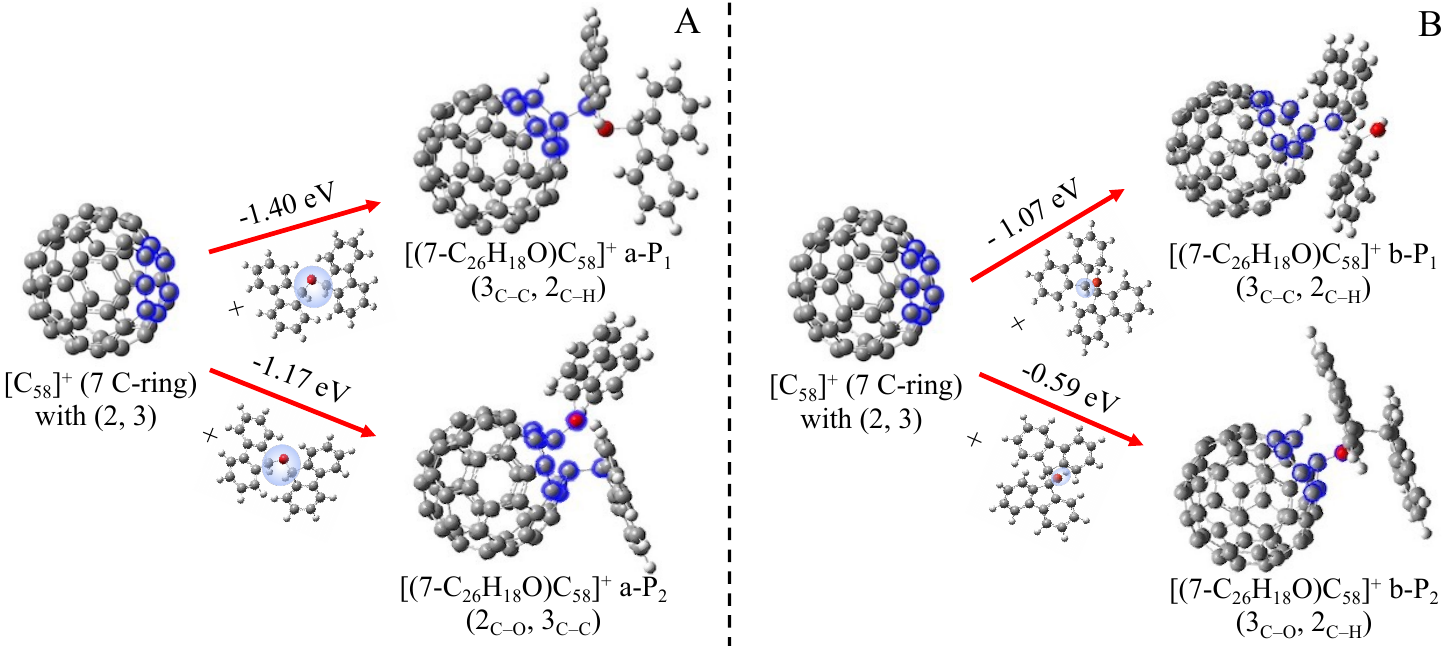}
	\caption{The Reaction pathways of C$_{58}$$^+$ (7 C-ring) with (2,3) $+$ C$_{26}$H$_{18}$O. The negative binding energy corresponds to the exothermic reaction. The 7 C-ring is highlighted.
	}
	\vspace*{-1.0em}
	\label{fig9}
\end{figure*}

Due to the high chemical reactivity shown in Fig.~\ref{fig6}, we selected C$_{58}$$^+$ (7 C-ring) with (2,3) for the formation of [(C$_{26}$H$_{18}$O)C$_{58}$]$^+$. The reaction pathways of C$_{58}$$^+$ (7 C-ring) $+$ C$_{26}$H$_{18}$O are illustrated in Fig.~\ref{fig9}, and we only provide reaction pathways related to the oxygenated functional groups (e.g., hydroxyl, OH and ether, C-O-C).

Fig.~\ref{fig9}(A) shows the reaction pathways related to the C-O-C group in C$_{26}$H$_{18}$O (P$_1$). [(7-C$_{26}$H$_{18}$O)C$_{58}$]$^+$ (a-P$_1$) is formed by H atom transfer reaction with exothermic energy of $-$1.40 eV ($-$32.3 kcal mol$^{-1}$), in which the C atom from the C-O-C group is added to the 7 C-ring of C$_{58}$$^+$ with the formation of a C-C single bond, and the H atom connected to the C-O-C group migrates to the adjacent carbon. [(7-C$_{26}$H$_{18}$O)C$_{58}$]$^+$ (a-P$_2$) is formed with the fracture of the C-O-C group. Releasing energy of $-$1.17 eV ($-$27.0 kcal mol$^{-1}$), two fragments after the fracture of the C-O-C group are connected to nearby carbon atoms on 7 C-ring of C$_{58}$$^+$. 

The reaction pathways related to the OH group in C$_{26}$H$_{18}$O (P$_2$) are selected and calculated in Fig.~\ref{fig9}(B). Since the exothermic energy of the OH group transfer reaction is close to that of the H atom transfer reaction on C$_{58}$$^+$ (7 C-ring), we only show two examples of H atom transfer reaction pathways. Accordingly, (7-C$_{26}$H$_{18}$O)C$_{58}$]$^+$ (b-P$_1$) is formed with the fracture of C-H bond near the OH group releasing energy of $-$1.07 eV ($-$24.7 kcal mol$^{-1}$), and [(7-C$_{26}$H$_{18}$O)C$_{58}$]$^+$ (b-P$_2$) is formed with the fracture of O-H bond releasing energy of $-$0.59 eV ($-$13.6 kcal mol$^{-1}$).

The exothermic energy for the formation of [(7-C$_{26}$H$_{18}$O)C$_{58}$]$^+$ is higher and can stabilize the whole clusters. Thus, we propose that [(C$_{26}$H$_{18}$O)C$_{58}$]$^+$ formed in the laboratory is a mixed cluster with different possible isomers. Interestingly, C$_{26}$H$_{18}$O is a large functionalized PAH molecule (with 45 atoms, and $\sim$ 2 nm in size) that can still react easily with fullerene species in the gas phase, which can prove again that fullerene species readily react with large molecules.

\section{Discussion}
\label{sec:dis}

During the collision reactions between fullerene cations and 9-hydroxyfluorene molecules, C$_{13}$H$_{10}$O and C$_{26}$H$_{18}$O are added to the surface of fullerene through sequential steps, resulting in a variety of fullerene/9-hydroxyfluorene cluster cations \citep{gar13b,sat13,boh16,zhen2019a,zhen2019b}. According to our calculations, the bonding ability, i.e., the reaction surfaces, reaction modes, and combination sites, play a decisive role in the cluster formation processes. 

To the reaction surfaces: two different reaction surfaces are considered--C$_{13}$H$_{10}$O ``landing'' on the 6 C-ring and C$_{13}$H$_{10}$O ``landing'' on the 7 C-ring, which is in agreement with previous works \citep{zhen2019b,hu23}. For 6 C-ring (both in C$_{60}$$^+$ and C$_{58}$$^+$), the stable product is the Van der Waals complex (with exothermic energy of $\sim$0.8 eV, 18.4 kcal mol$^{-1}$); for 7 C-ring, 9-hydroxyfluorene tends to form covalently bonded clusters and connect directly to the 7 C-ring surfaces of C$_{58}$$^+$ (with exothermic energy of $\sim$1.1 eV, 25.4 kcal mol$^{-1}$). The exothermic energy of the covalently bonded cluster formed on the 7 C-ring surfaces is higher than that on the 6 C-ring surfaces. Therefore, the enhanced chemical reactivity for smaller fullerene cations is mainly attributed to the deformed rings (e.g., 7 C-ring in C$_{58}$$^+$) \citep[e.g.][]{zhen2019b,hu21b,hu23}.

To the reaction modes: five different reaction modes are taken into consideration--the Van der Waals interaction \citep[e.g.][]{zhen2019b,hu23}, the H transfer reaction from CH unit, the H transfer reaction from OH group, the OH group transfer reaction, and the reaction concerning the aromatic ring with the newly formed double C-C bonds \citep{pet93,sat13,boh16,zhen2019b}. We can see that the reaction concerning the aromatic ring is endothermic, i.e., fullerenes and 9-hydroxy fluorene molecules are difficult to form the cluster connected by double C-C bonds \citep{zhen2019a,zhen2019b}.

To the combination sites: due to the geometric structure of 7 C-ring, four combination modes are considered for the OH group transfer reactions, (1,2)=(1,2$'$), (2,3)=(2$'$,3$'$), (3,4)=(3$'$,4$'$), (4,4$'$). By comparing the binding energies for different combination modes, the chemical reactivity can be obtained as follows: (2,3)=(2$'$,3$'$)$\sim$(3,4)=(3$'$,4$'$)$>$(4,4$'$)$>$(1,2)=(1,2$'$). Thus, we can conclude that the formation of C$_{58}$$^+$/9-hydroxyfluorene cluster tends to occur along the boundary of the 7 C-ring and 5 C-ring rather than the boundary of the 7 C-ring and 6 C-ring, which suggests that the combination sites adjacent to 5 C-rings may have a higher chemical reactivity \citep{pet93}.

Overall, we infer that during the formation of fullerene/9-hydroxyfluorene cluster cations, the first C$_{13}$H$_{10}$O molecule may initially be connected to the deformed cage surfaces of smaller fullerene cations. After the deformed rings are occupied, other C$_{13}$H$_{10}$O molecules will be added to the 6 C-ring surfaces by Van der Waals bonds. As the number of C atoms decreases, there may be more deformed rings in smaller fullerene cations \citep[e.g.][]{hu21b}. Therefore, more C$_{13}$H$_{10}$O molecules can be added to the cage surfaces of smaller fullerene cations, resulting in various large fullerene/9-hydroxyfluorene cluster cations \citep{zhen2019a,zhen2019b,hu21a}. 

\section{Astronomical implications}
\label{sec:astro}

\begin{figure*}
	\centering
	\includegraphics[width=\textwidth]{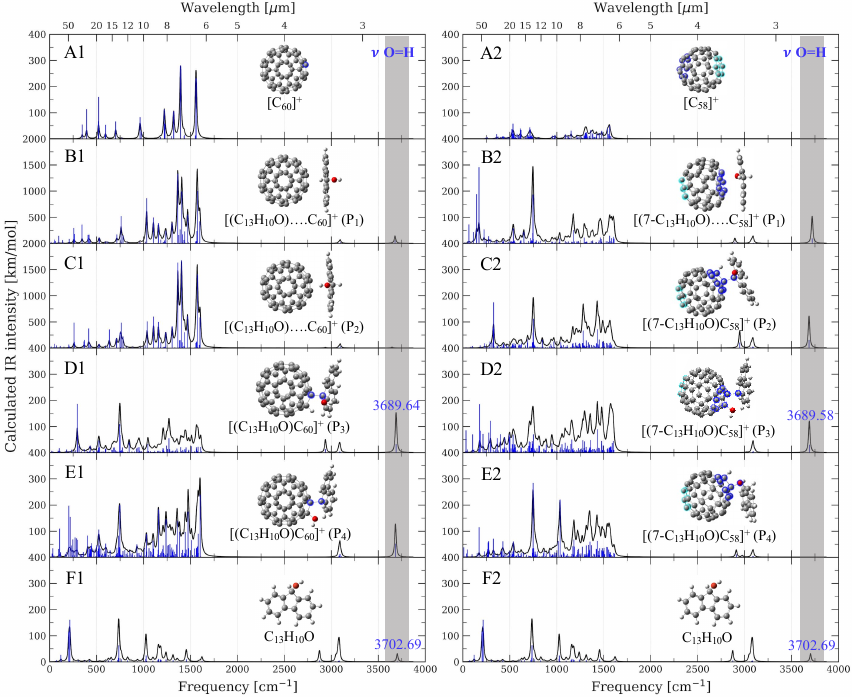}
	\vspace*{-2.0em}
	\caption{Computed IR spectra and vibrational normal modes for fullerene/9-hydroxyfluorene cluster cations: panels (A1$-$F1) are for C$_{60}$$^+$ $+$ C$_{13}$H$_{10}$O, and panels (A2$-$F2) for C$_{58}$$^+$ $+$ C$_{13}$H$_{10}$O, respectively. The vibrational band positions were scaled by 0.967 for C-H stretch mode and 0.980 for all other modes.
	}
	\vspace*{-1.0em}
	\label{fig10}
\end{figure*}

\begin{figure*}
	\centering
	\includegraphics[width=\textwidth]{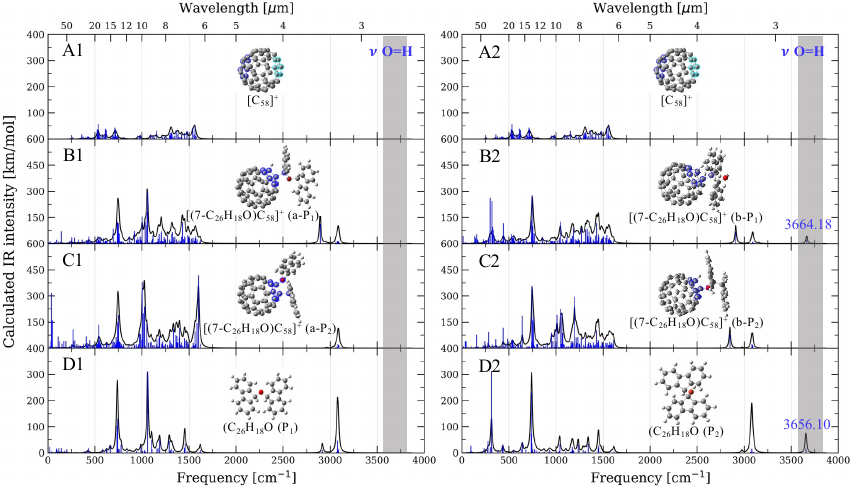}
	\vspace*{-2.0em}
	\caption{Computed IR spectra and vibrational normal modes for C$_{58}$$^+$/C$_{26}$H$_{18}$O cluster cations: panels (A1$-$D1) are for C$_{58}$$^+$ $+$ C$_{26}$H$_{18}$O (P$_1$), and panels (A2$-$D2) for C$_{58}$$^+$ $+$ C$_{26}$H$_{18}$O (P$_2$), respectively. The vibrational band positions were scaled by 0.967 for C-H stretch mode and 0.980 for all other modes.
	}
	\label{fig11}
	\vspace*{-1.0em}
\end{figure*}

In interstellar environment, the forms and states of fullerene molecules are affected and constrained by surrounding environmental factors, such as interstellar UV radiation, the flux of H atoms, hydrogen ions, and other coexisting molecules \citep{gar10,gar13b,dun13,omo16}. These environmental factors provide feedback and constraints on the states and forms of these fullerene molecules at their different evolution periods. Based on the interstellar coevolution network, in this work, the fullerene molecules under the environmental factors of coexisting oxygenated functional substituted PAH molecules are studied through experiments and theoretical quantum calculations.

The experimental results show that the reactions between fullerene cations (C$_n$$^+$, $n$$=$32, 34, ..., 60) and neutral 9-hydroxyfluorene molecules can occur in the gas phase. Fullerene/9-hydroxyfluorene cluster cations are formed through a sequential-step pathway, in which 9-hydroxyfluorene molecules are added repeatedly onto the cage surface of fullerene cations \citep{gar13b,sat13,zhen2019a}. With the number of C atoms decreasing, more C$_{13}$H$_{10}$O molecules can be added to smaller fullerene cations, suggesting an enhanced chemical reactivity for these smaller fullerene cations. The higher reactivity of small fullerenes may be attributed to the diversity of molecular structures \citep{pet93,boh16,zhen2019b,hu21b} generated by UV laser photolysis, including caged, semi-caged, or planar structures \citep{ber12,can19}. 9-hydroxyfluorene molecules have a relatively low chemical reactivity in the reaction with fullerene cations compared with other substituted derivative PAHs, such as anthracene, 9-methylanthracene, 9-vinylanthracene and 9-aminoanthracene \citep{zhen2019a,zhen2019b,hu21a}, indicating the effect of oxygenated functional group on the chemical reactivity. 

From the results of the theoretical calculation, the molecular structure of fullerene/9-hydroxyfluorene cluster cations are diverse. The Van der Waals complexes are stable and may be readily formed when C$_{13}$H$_{10}$O molecules are added to the 6 C-ring surfaces. In the reactions on the 7 C-ring surfaces, 9-hydroxyfluorene tends to be covalently bonded with fullerene cations. The higher exothermic energy for 7 C-ring surfaces also suggests an enhanced chemical reactivity of smaller fullerene cations \citep[e.g.][]{zhen2019b,hu21b,hu23}, which is well consistent with the experimental results. The relative chemical reactivity of these species is very important in the evolution network of the ion-molecular reaction, which affects the abundance ratio of formed species \citep{jag09,mon13,zhen2019b}. 

Since the fullerenes (e.g., C$_{60}$ and C$_{60}$$^+$) have been confirmed in the ISM, we expect their derivatives (i.e., fullerene/PAH clusters) might also coexist in space. However, due to the non-overlapping spatial distributions of fullerenes and smaller PAHs in planetary nebulae, we do not propose fullerene/PAH adducts as contributors. Especially in the harsh interstellar environment, small PAHs and their derivates hardly survived \citep{tie08}. Nevertheless, species formed by reactions between fullerene cations and molecules found in planetary nebulae may be involved \citep{gar10,cam10}. The coexisting molecules, not only specific PAHs but also mixed PAHs containing oxygenated functional groups (OH or C-O-C), may be involved in the growth of dust particles. 

In addition, we also obtained the dehydrated product C$_{26}$H$_{18}$O in the heat processes, which is a large functionalized PAH molecule (with 45 atoms, and in $\sim$ 2 nm in size). From the obtained results, C$_{26}$H$_{18}$O can react readily with fullerene species in the gas phase, supporting the reaction of fullerene species with large molecules during the formation of dust particles.

The newly formed fullerene/9-hydroxyfluorene cationic clusters may also be candidates for the observed IR interstellar bands. They may exist in the same interstellar area through an evolutionary ion-molecular reaction network between substituted PAHs and fullerenes. This may contribute to the observed interstellar spectrum and motivate spectroscopic studies \citep{tie13}. 

Fig.~\ref{fig10} demonstrates the calculated IR spectra for some possible typical fullerene/9-hydroxyfluorene clusters: panels (A1 \& A2) are for C$_{60}$$^+$ and C$_{58}$$^+$, respectively; panels (B1-E1) for [(C$_{13}$H$_{10}$O)C$_{60}$]$^+$ (P$_1$-P$_4$); panels (B2-E2) for [(7-C$_{13}$H$_{10}$O)C$_{58}$]$^+$ (P$_1$-P$_4$); panels (F1 \& F2) for C$_{13}$H$_{10}$O. Similarly, Fig.~\ref{fig11} demonstrates the calculated IR spectra for some possible typical C$_{58}$$^+$/C$_{26}$H$_{18}$O clusters: panels (A1 \& A2) are for C$_{58}$$^+$; panels (B1$\ \&\ $C1) for [(C$_{26}$H$_{18}$O)C$_{58}$]$^+$ (a-P$_1$ \& a-P$_2$); panels (B2$\ \&\ $C2) for [(C$_{26}$H$_{18}$O)C$_{58}$]$^+$ (b-P$_1$\ \&\ b-P$_2$); panels (D1 \& D2) for C$_{26}$H$_{18}$O (P$_1$ \& P$_2$). The vibrational band positions were scaled by a mode-dependent scaling factor, which is 0.967 for C-H stretch mode and 0.980 for all other modes \citep{bau97,boe14}, and the black line is the spectrum simulated by Gaussians with a full-width at half-maximum of 10 cm$^{-1}$. 

As shown in Fig.~\ref{fig10} and Fig.~\ref{fig11}, the IR spectra are very complex, in which many vibration modes are difficult to identify. In general, the IR spectra of fullerene/9-hydroxyfluorene derivate cluster cations were different from that of fullerene cations and C$_{13}$H$_{10}$O or C$_{26}$H$_{18}$O. There are some new vibrational modes, some missing ones, and some reserved ones. The IR spectra with different binding modes are different in the peak positions and IR intensities. Different cage surfaces also contribute to different IR spectra, since the intensities of each vibrational peak are different in the spectra of C$_{58}$$^+$/9-hydroxyfluorene clusters on the 7 C-ring surfaces or C$_{60}$$^+$/9-hydroxyfluorene clusters on the 6 C-ring surfaces. 

We only conclude one characteristic peak in the range of 3500-3800 cm$^{-1}$: the O-H stretch vibration. The O-H stretch features have been observed in space, especially in interstellar ices \citep[e.g.,][]{all92, tie96}. Since the cationic clusters in this study are relevant to planetary nebulae devoid of such ices, the O-H stretch vibrational mode in observations of planetary nebulae may be a reliable indicator of fullerene/9-hydroxyfluorene or similar clusters. The peak positions and intensities of the O-H stretch vibration change with different binding modes, which is shown in Fig.~\ref{fig10}(B1-F1 \& B2-F2) and Fig.~\ref{fig11}(B1-D1 \& B2-D2). For example, in Fig.~\ref{fig10}(D1 \& D2), with regard to the IR spectrum of [(C$_{13}$H$_{10}$O)C$_{60}$]$^+$ (P$_3$) and [(C$_{13}$H$_{10}$O)C$_{58}$]$^+$ (P$_3$), the characteristic peaks are 3689.64 and 3689.58 cm$^{-1}$, respectively; in Fig.~\ref{fig10}(F1 \& F2), concerning the IR spectrum of C$_{13}$H$_{10}$O, the characteristic peak of the O-H stretch vibration is 3702.69 cm$^{-1}$. We can see that for the clusters formed with the fracture of the O-H bond (e.g., [(C$_{13}$H$_{10}$O)C$_{58}$]$^+$ (P$_4$), [(C$_{26}$H$_{18}$O)C$_{58}$]$^+$ (a-P$_1$, a-P$_2$ \& b-P$_2$)), the O-H stretch vibration mode is missing. Interestingly, as for Van der Waals molecules with different alignments ([(C$_{13}$H$_{10}$O)....C$_{60}$]$^+$ (P$_1$ \& P$_2$)), there is an obvious difference in the peak position and intensity of the O-H stretch vibration, while the peak position and intensity of other vibration modes are almost the same.

Overall, these calculated IR spectra reveal the complexity of fullerene/9-hydroxyfluorene cluster cations, which suggests the significant role of different geometric structures, including alignment, cage surface, and connection type. Due to the structure of fullerene/9-hydroxyfluorene clusters initially formed being diverse, for deeply understanding the evolution of fullerene/PAH derivatives in cosmic environment, further observational and theoretical studies are necessary, such as more high-resolution observation data (e.g., the James Webb Space Telescope) or extensive theoretical calculations involving fullerenes \citep{ber22,bar23,lai23,xu23,yang23}. 

\section{Conclusions}
\label{sec:con}

Combining experiments with quantum chemical calculations, the reaction processes of fullerene cations with oxygenated functional PAH molecules are studied in this work. Fullerene/9-hydroxyfluorene cluster cations are efficiently formed through an ion-molecule collision reaction in the gas phase. Through theoretical quantum calculations, the molecular structures of newly formed clusters (e.g., [(C$_{13}$H$_{10}$O)C$_{60}$]$^+$, [(C$_{13}$H$_{10}$O)$_{1-2}$C$_{58}$]$^+$, and [(C$_{26}$H$_{18}$O)C$_{58}$]$^+$), the binding energies of their reaction pathways, together with their IR spectra, are obtained. Smaller fullerene cations have an enhanced chemical reactivity due to their deformed carbon rings (e.g., 7 C-ring). The bonding ability plays a decisive role in the cluster formation processes. Various factors that affect chemical reactivity differently are discussed, such as reaction surfaces, reaction modes, and combination sites. 

The results we obtained once again validate the complexity of interstellar molecules and the diversity of the evolution for fullerene species under the constraints of coexisting PAH molecules. Hence, if these fullerenes are present in space, forming fullerene-based clusters could produce an extended family of large molecules. In addition, the results we obtained provide insights into interstellar chemistry of fullerene species and the geometry structures and functional groups that may affect the formation process of fullerene-PAH-derived cluster cations in the ISM.

\section*{Acknowledgements}

This work is supported by the National Natural Science Foundation of China (NSFC, Grant No. 12333005, 12122302, and  12073027). Theoretical calculations were performed at the Supercomputing Center of the University of Science and Technology of China. 

\section*{Data availability}

All data generated or analyzed during this study are included in this article.

\appendix

\section{Experimental methods}

The experiment was performed in the setup equipped with the quadrupole ion trap and a reflection time-of-flight (QIT-TOF) mass spectrometry. More detailed information is provided in \cite{zhen2019b}. Briefly, the fullerene molecule (C$_{60}$) was sublimated in an oven at a temperature of $\sim$ 613 K and ionized by an electron gun (Jordan, C-950, $\sim$ 82 eV). The fullerene cations of interest were then transported into the quadrupole ion trap (Jordan, C-1251) via an ion gate and a quadrupole mass filter (Ardara, Quad-925mm-01).

The gas-phase 9-hydroxyfluorene molecules were produced by heating their powder (J\&K Scientific, with purity better than 99 \%) in another oven (heating at $\sim$ 380 K) mounted over the ion trap, which was effused continuously towards the center of the ion trap during the experiment. In the ion trap, fullerene/9-hydroxyfluorene cluster cations were formed through ion-molecule collision reactions between fullerene cations and 9-hydroxyfluorene molecules. Buffer gas helium with high purity was used to thermalize the ions through collision ($\sim$ 300 K). The pressures in the ion trap chamber were $\sim$ 6.0 $\times$ 10$^{-7}$ and $\sim$ 2.0 $\times$ 10$^{-7}$ mbar in the cases with and without adding 9-hydroxyfluorene molecules, respectively.

To generate a larger amount of photo-fragments of small fullerene ions, a third harmonic output (355 nm) of a Nd:YAG laser (Spectra-Physics, INDI, pulse width $\sim$ 6 ns, frequency 10 Hz) was used to irradiate the trapped fullerene cations. Furthermore, to obtain detailed formation pathways of each fullerene cation, the stored waveform inverse Fourier transform excitation (SWIFT) isolation technique \citep{dor96} was introduced to the ion trap that allows to select the desired individual fullerene cations with specific mass/charge ($m/z$) range.

The whole set of experiments was conducted at a typically measured frequency of 0.1 Hz, i.e. each measurement period is 10 s. A high precision digital delay/pulse generator (SRS, DG535) was used to control the time sequence, which was set as follows:\\ (1) Without SWIFT isolation: the ion gate kept open during the time interval 0$-$4.0 s, allowing C$_{60}$$^+$ to accumulate to a certain amount, and the beam shutter then kept open during 4.0$-$5.8 s, allowing trapped ions to be irradiated by a 355 nm laser beam. After the irradiation, the newly formed fullerene ions further reacted with 9-hydroxyfluorene molecules for several seconds. At 9.88 s, the ions were introduced out of the ion trap and then into a mass spectrometer to be detected; \\(2) With SWIFT isolation: the ion gate is kept open during the time interval 0$-$4.0 s, and the beam shutter is then kept open during 4.0$-$5.8 s. After the irradiation, a SWIFT isolation technique was employed to select the fullerene cations of interest (5.8$-$6.0 s). The selected ions then react with 9-hydroxyfluorene molecules for 3.88 s (6.0$-$9.88 s). At 9.88 s, the ions were introduced out of the ion trap and then into a mass spectrometer to be detected.

\section{Charge Population Analysis}

In Section~\ref{sub:theo}, the formation of two different Van der Waals molecules [(C$_{13}$H$_{10}$O)....C$_{60}$]$^+$ (P$_1$) and (P{$_2$}) have almost the same exothermic energy of $-$0.82 eV ($-$18.9 kcal mol$^{-1}$). Here, we provided a mulliken charge analysis performed by the Gaussian 16 program \citep{fri16} package, using density functional theory employing the hybrid functional B3LYP \citep{bec92,lee88} with the 6-311++G(d,p) basis set. We also visualized the HOMO and LUMO of these Van der Waals molecules, respectively.

\begin{figure}
	\centering
	\includegraphics[width=3.0in]{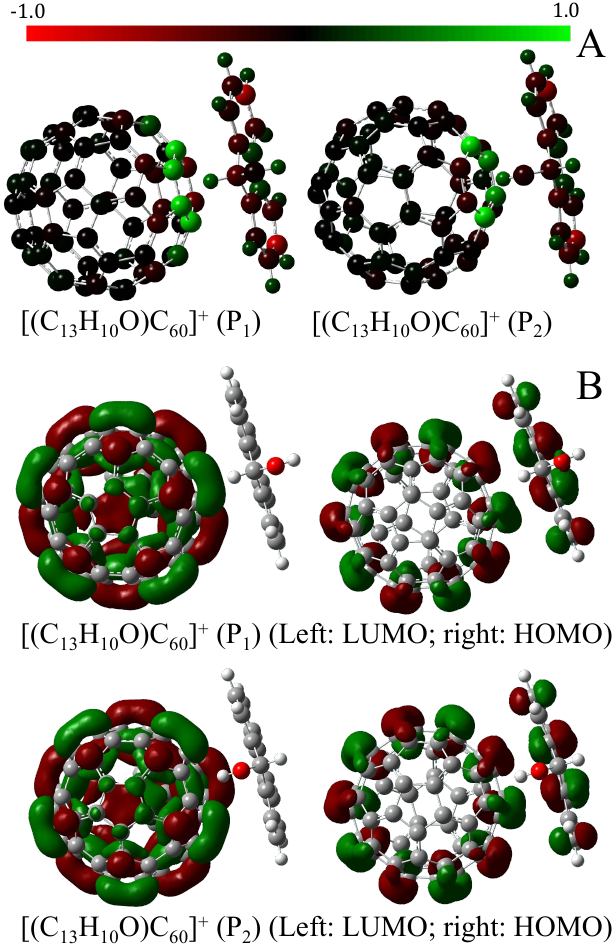}
	\vspace{-0.5em}
	\caption{The 3D images of the Mulliken charge separation (A), the HOMO and LUMO (B) for Van der Waals molecules [(C$_{13}$H$_{10}$O)....C$_{60}$]$^+$ (P$_1$) and (P{$_2$}). In the Mulliken charge separation, atoms are colored according to charge. Red and green correspond to negative and positive electricity, respectively.
	}
	%\vspace{-0.5em}
	\label{figB1}
\end{figure}

Fig.~\ref{figB1} shows the 3D images of the Mulliken charge separation, the HOMO and LUMO for these two Van der Waals molecules [(C$_{13}$H$_{10}$O)....C$_{60}$]$^+$ (P$_1$) and (P{$_2$}), respectively. As demonstrated in Fig.~\ref{figB1}, [(C$_{13}$H$_{10}$O)....C$_{60}$]$^+$ (P$_1$) and (P{$_2$}) have similar charge separation and orbital behavior, which implies similar stability of different Van der Waals molecules. During the formation of Van der Waals molecules, the cage surfaces of C$_{60}$$^+$ and the aromatic rings of C$_{13}$H$_{10}$O are dominant and may be related to binding energy. We can see that there is only a small difference in electronegativity of O atoms between the two molecules. Therefore, we think the almost similar exothermic energy may be reliable, and different alignment may not affect significantly.

\end{document}